\documentclass[twocolumn,secnumarabic,amssymb, nobibnotes, aps, prb]{revtex4-1}

\usepackage{graphicx}
\usepackage{dcolumn}
\usepackage{bm}
\usepackage{hyperref}
\usepackage{CJK}

\begin{document}


\title{ Nonlinear magnetic metamaterials and possible applications on all-optical comparators and bistabilities in Fabry-Perot cavities}

\author{Yi S. Ding}
\email{dyi@pku.edu.cn}
\author{Ruo-Peng Wang}%
 \email{rpwang@pku.edu.cn}
\affiliation{%
 State Key Laboratory for Mesoscopic Physics, Department of Physics, Peking University,
Beijing 100871, People's Republic of China
}%




\date{\today}

\begin{abstract}
We investigate the modulational instability and time-domain dynamics of nonlinear magnetic metamaterials composed of coupled split-ring resonators loaded by Kerr nonlinearity. Our results indicate that the recently proposed optical switching of local optical index based on uniform-response assumption seems fragile. We conceive two alternative schemes to utilize the valuable enhanced nonlinearity, one is to focus on few-body systems and directly make use of the modulational instability (e.g., an optical comparator design), the other is to consider global switching arising from global feedbacks as in usual cases (e.g., Fabry-Perot cavities) rather than local-resonances-based switching of optical constants which may be destroyed by the discreteness and interactions as we show in this paper. We also try to provide comprehensive understanding of the relations between our results on discrete nonlinear metamterials and those on continuous metamaterials and natural materials in the literatures.

\begin{description}
\item[PACS numbers]
\verb| 42.65.Pc, 42.65.An, 75.30.Kz, 64.60.F-|
\end{description}
\end{abstract}

\maketitle

\section{introduction}\label{int}


Linear metamaterials, ever since the theoretical works \cite{pendry1999,pendry1996,walser2001,pendry2000} and the experimental works \cite{smith2000,shelby2001}, have been highlighted by a great number of achievements. Nonlinear metamaterials, however, as Zheludev points out in [\onlinecite{zheludev2010}], are sill ``green apples'' on the knowledge tree for metamaterials.

In the discussions of nonlinear metamaterials, particularly those composed of nonlinear split-ring resonators (SRR), it will be of great convenience to treat the metamaterials as continuous materials with controllable optical constants [e.g. $\mu(I)$ and $n(I)$], which allows us to employ existing concepts and methods in nonlinear optics. Several works since [\onlinecite{zharov2003}], including [\onlinecite{lapine2003,agranovich2004,wen2006,joseph2010,roppo2010,zharova2005}], focus on this approach. They either directly assume intensity-dependent-index continuous media or assume uniform responses of the discrete system before calculating optical constants. These approaches share a common implicit assumption, i.e., on a scale much smaller than the wavelength, a cluster of discrete inclusions in a metamaterial should uniformly (at least slowly-varyingly) respond to external radiation. This assumption justifies the continuous descriptions which investigate the materials by partial differential equations. Particularly, some of them ([\onlinecite{zharov2003}] and [\onlinecite{chen2011}]) propose applications for global-geometry-independent optical switching based on the bistable properties of these local controllable optical constants. Note that these bistabilities  \cite{zharov2003,chen2011} are obtained after assuming that all the meta-atoms respond uniformly in the materials, at least in the equal phase planes, just as in the linear cases.

However, we should note, that nonlinear metamaterials (usually composed of nonlinear SRR) are intrinsically discrete, and that they can even present strong local feed-back at the ``single-atom'' level, unlike the conventional nonlinear optical materials to which the feed-back is usually introduced globally (e.g. by two mirrors in a Fabry-Perot cavity). Therefore, we expect to gain more knowledge if we treat the nonlinear metamaterials as discrete, interacting nonlinear lattices as they are.

Some results have been obtained by treating the nonlinear metamaterials as discrete nonlinear lattices, including [\onlinecite{lazarides2006,eleftheriou2008,cui2009,eleftheriou2009,lazarides2010,shadrivov2006,rosanov2012}]. What we should pay particular attentions is the series of works \cite{lazarides2006,eleftheriou2008,eleftheriou2009,rosanov2012}, in which the authors show that the steady responses of dissipative SRR arrays may violate the translational symmetry and can be in the multibreather states \cite{flach2008} even when the arrays are driven by uniform radiation. The breather states in these works are theoretically constructed in the week-interaction regime and in the anti-continuous limit \cite{aubry1997} in which the states are \emph{a priori} nonuniform and may in practice be excited by introducing chirping radiation \cite{lazarides2009,lazarides2009pre}.

In this paper, by investigating nonlinear effective circuit models, we show that even when we start from the nearly-zero states under uniform external radiations, modulational instabilities will step in the bistable ranges proposed in [\onlinecite{zharov2003}], i.e., the responses of the SRR array will spontaneously break the translational symmetry , and finally arrive at nonuniform states or even be chaotic, which will make the definition of optical constants in [\onlinecite{zharov2003}] invalid at least in the bistable range of interest. These results reveal that the switching of field-dependent optical constants may not exist due to modulational instabilities. Despite the negative results, we propose the possibility of constructing all-optical comparators based on the spontaneous-symmetry-breaking phenomenon of two strongly coupled SRR. We also propose to utilize the valuable enhanced nonlinearity in SRR arrays, by introducing global feed-back mechanism as in usual cases. Finally, we comment on the relations of the results here based on our discrete models for metamaterials and those in the literatures.

\section{The theoretical model}\label{single}

When a plane wave is normally incident into a three-dimensional nonlinear metamaterial composed of SRR, if we adopt the uniform assumption as in \cite{zharov2003}, the SRR located on an equal-phase plane of the wave will experience uniform external driving force (induced by other SRR and the external radiation) and respond with uniform amplitude and phase. We can thus simplify our considerations by focusing on a two-dimension (or even one-dimension) array of SRR radiated by uniform external fields.  If we find that in some cases the uniform assumption is not valid even in the equal phase plane, we can undoubtedly conclude that such assumption is not valid in the three-dimension bulk materials.

We model a rectangle lattice of SRR by LCR circuits coupled by mutual inductance (just as in \onlinecite{zharov2003}) with nearest-neighbor approximations and periodic boundary conditions. In terms of dimensionless quantities, the dynamic equation of the lattice under slowly-varying approximation is

\begin{eqnarray}
&&(2i\Omega+\gamma)\frac{dq_{i,j}}{d\tau}+2i\Omega\kappa_x\left(\frac{dq_{i-1,j}}{d\tau}+\frac{dq_{i+1,j}}{d\tau}\right)\nonumber\\
&&+2i\Omega\kappa_y\left(\frac{dq_{i,j-1}}{d\tau}+\frac{dq_{i,j+1}}{d\tau}\right)=u_{i,j}+\nonumber\\
&&\kappa_x\Omega^2(q_{i-1,j}+q_{i+1,j})+\kappa_y\Omega^2(q_{i,j-1}+q_{i,j+1})\nonumber\\
&&-(-\Omega^2+i\gamma\Omega+1-|q_{i,j}|^2)q_{i,j}\nonumber\\\label{dynamic}
\end{eqnarray}

where $u_{i,j}$, $q_{i,j}$ are the (dimensionless) local electro-motive force acting on the SRR at site $(i, \,\,j)$ and the electric charge hold by its capacitor, respectively. The $\Omega$ is the dimensionless frequency; $\tau$, the time; $\gamma$, the resistance; $\kappa_{x,y}$, mutual inductances between nearest neighbors in the `x, y' directions, respectively. Then the steady states should satisfy the equation
\begin{eqnarray}
&&u_{i,j}+\kappa_x\Omega^2(q_{i-1,j}+q_{i+1,j})+\kappa_y\Omega^2(q_{i,j-1}+q_{i,j+1})\nonumber\\
&&-(-\Omega^2+i\gamma\Omega+1-|q|^2)q_{i,j}=0\label{steady}
\end{eqnarray}

More detailed definitions of the quantities are presented in Appendix \ref{A}.

\section{Modulational instabilities of uniform responses of an SRR array}
The modulational instabilities of continuous nonlinear materials with negative refraction index have been studied in [\onlinecite{wen2006}], in which temporal, spatial, and temporal-spatial  modulational instabilities are all studied. The spatial modulational instabilities of three-dimension discrete nonlinear SRR arrays induced by magneto-inductive waves have been studied in [\onlinecite{shadrivov2006}], in which it is shown that the unstable range may completely cover the decreasing range of the bistability, especially the two switching points. In this section, we in general employ the formalism in [\onlinecite{shadrivov2006}] but focus on the model (\ref{dynamic}) introduced in the previous section.

Under uniform-response assumption as in [\onlinecite{zharov2003}], i.e., $u_{i,j}=u$ and $q_{i,j}=q$, the Eqs.(\ref{steady}) for steady states are simplified
\begin{equation}
u+(2\kappa_x\Omega^2+2\kappa_y\Omega^2+\Omega^2-i\gamma\Omega-1+|q|^2)q=0.\label{uniform-steady}
\end{equation}
We can solve this equation and obtain the bistabilities of uniform responses as illustrated in Fig.(1). However, since mutual interactions in densely stacked metamaterials are inevitably present, it is a nature question that whether possible excitations of magneto-inductive waves (spatial Fourier components of fluctuations) could induce modulational instabilities of these uniform states. By linear stability analysis (e.g.,[\onlinecite{flach2008,shadrivov2006}]), we can obtain the growth rate for each Fourier component of magneto-inductive waves Eq.(\ref{growrate}).
If for all Fourier components, $\lambda(\mathbf{k})$ has a negative real part, the corresponding state is stable. Otherwise, unstable.

\begin{figure}[h]\centering
\includegraphics[height=7cm,width=7cm]{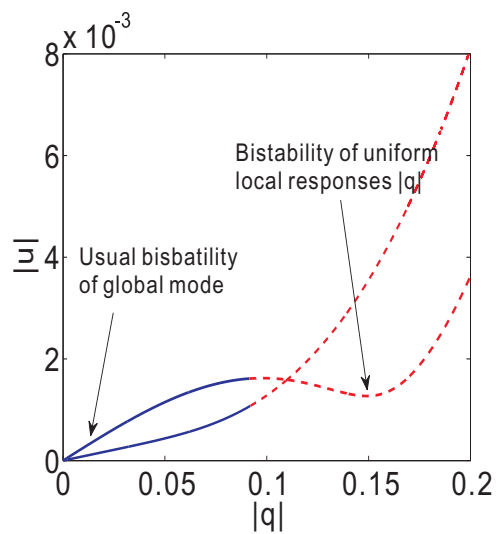}
\caption{The stability analysis of uniform response. For the curve with uniform bistability, $\Omega=0.835$; for the other, $\Omega=0.850$. Other parameters: $\kappa_x=0.3$, $\kappa_y=-0.1$, $\gamma=0.01$.}\label{unif}
\end{figure}

After the linear stability analysis, the stable and instable uniform responses are denoted by solid and dashed segments in Fig.(\ref{unif}), respectively.
Since we have chosen an equal phase plane to investigate, the mutual inductances between nearest neighbors should be positive in one direction and negative in the other direction. In all cases in Fig.(\ref{unif}), we can clearly identify critical points as the boundaries between the stable and instable ranges. As the system is tuned from the stable ranges (across the critical points) to the instable ranges, we expect that it will spontaneously break the translational symmetry and actually be in nonuniform states. We can also observe from Fig.(\ref{unif}) and Eq.(\ref{growrate}) that in the low intensity limit, the modulational instability is absent for all cases implying that nonlinear metamaterials always allow uniform description in perturbative regimes.

\section{Time-domain response of a two-dimension SRR array}

\begin{figure}[h]\centering
\includegraphics[height=15cm,width=8cm]{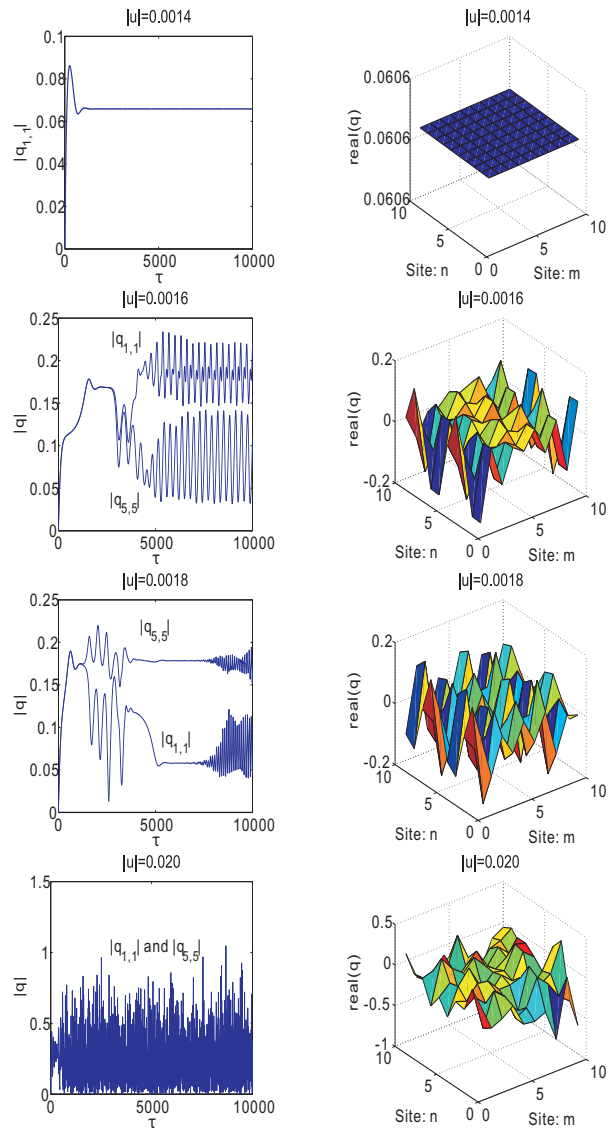}
\caption{The time-domain dynamics for different radiation intensities. For each intensity the final states after $\tau=10^4$ time of evolving are plotted in 3D figures.}\label{complex}
\end{figure}

A natural question arises that to what degree the actual response can deviate from the uniform assumption. In order to show an explicit picture of typical nonuniform responses when a system has evolved into the unstable range, we directly integrate Eqs.(\ref{dynamic}) using the Runge-Kutta method. In the recent papers \cite{lazarides2006,eleftheriou2008,eleftheriou2009}, single breather states (localized states which break the translational symmetry) are constructed from the anti-continuous limit in the weak-coupling regime. But these methods possibly can not be used in our problems since the coupling may be too strong for a single breather's existence. Here, we start from nearly-zero states with tiny random fluctuations and turn on the uniform radiation at the beginning. We then monitor the time-domain responses of two-dimension model (containing 10 $\times$ 10 SRR) under different intensities of radiation. The results are shown in Fig.(\ref{complex}). The parameters are chosen as the case with uniform bistable response in Fig.(\ref{unif}). In that case, the critical radiation intensity $u$ is about $0.0016$ which is the boundary between the stable range and unstable range. In Fig.(\ref{complex}), we test four radiation intensities $|u|=0.0014$, $0.0016$, $0.0018$ and $0.020$, of which the former three are near the critical value while the fourth is deep in the unstable range. The initial response is nearly zero with each individual response randomly distributing within $[0, 10^{-6}]$. For each test intensity, we evolve the system for $10^4$ periods and plot the time-domain responses of the the SRR at two different sites, (1,1) and (5,5). The final states for each case are also plotted in 3D figures. We observe that for the intensity below the critical value $|u|=0.0014$, the response of the system quickly converge to a steady state which is uniform for all SRR, which means that the uniform assumption in [\onlinecite{zharov2003}] is justified. However, when the intensity enters the unstable range, e.g., $|u|=0.0018$, the system spontaneously evolve into nonuniform states and cannot even reach a steady state (it seems like a limit circle state with periodic oscillation). For the case very near to the boundary $|u|=0.0016$, the situation is similar to $|u|=0.0018$ with the translational symmetry also drastically broken. When we come to the case in the deep unstable range $|u|=0.020$, it is even worse is that the time-domain responses quickly become chaotic. A common characteristic for the cases with nonuniform responses is that the actual states are highly disordered with individual responses being randomly positive and negative, which can hardly support regular wave propagation.

\section{Optical comparators based on the spontaneous symmetry breaking}
We can see from the above two sections that the responses of nonlinear metamaterials in the unstable ranges can be rather complex. If we want to make use of these nonuniform states, apart from directly investigating the details of the complexity (e.g., [\onlinecite{cui2009,eleftheriou2008}]) in these ranges, we may also focus on few-body systems to avoid these complexities.

\begin{figure}[h]\centering
\includegraphics[height=4cm,width=8cm]{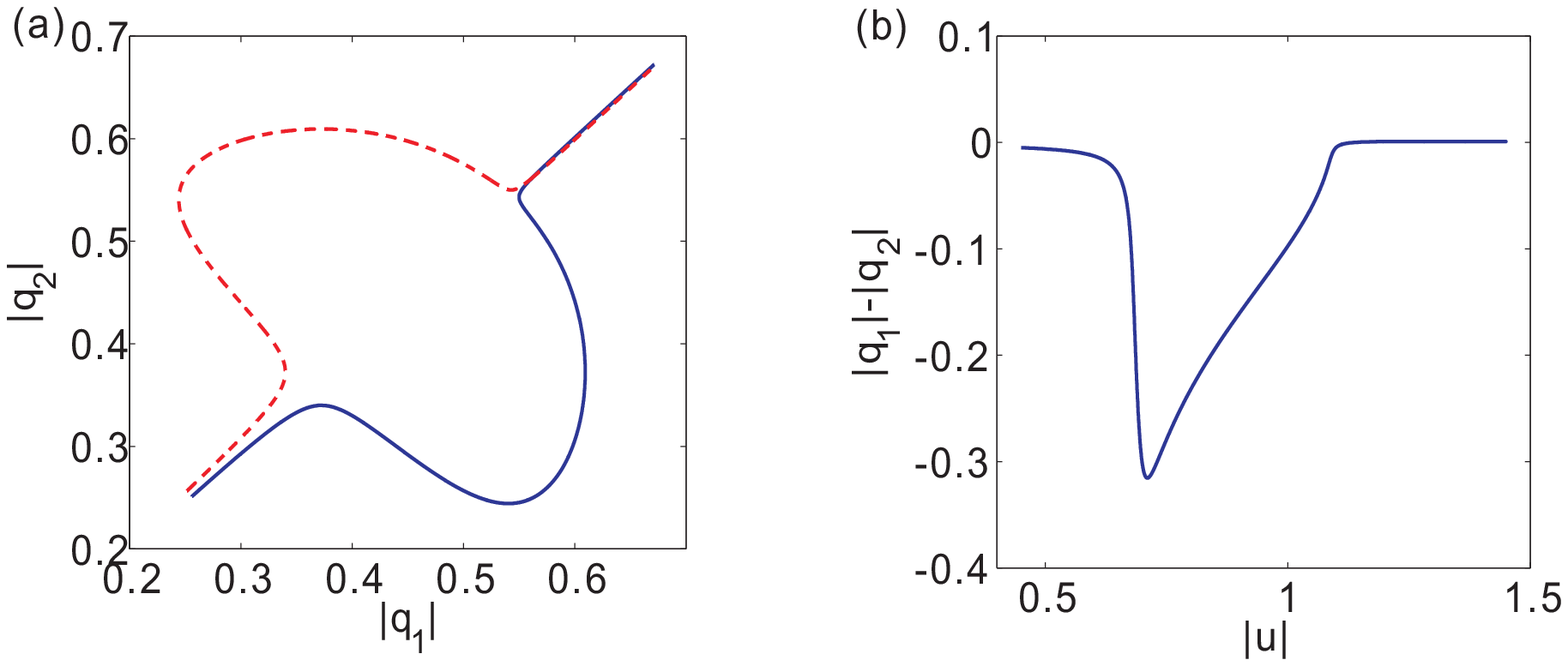}
\caption{a): The principle of the proposed comparator made of two strongly coupled identical SRR. The solid (dashed) line is for $u_1-u_2=0.001(-0.001)$, respectively. b): The deviation from uniform states as functions of overall radiation intensity. The stability of the uniform states can be guaranteed because of the continuation procedure. Parameters: $\Omega=1.3$, $\gamma=0.1$, $\kappa=0.6$.}\label{comparator}
\end{figure}

The spontaneous symmetry breaking (bifurcation) and its sensitivity on small deviations indicate that this phenomenon can be utilized to construct comparators. In Fig.\ref{comparator}, we illustrate this possible application by plotting the evolution of the steady states (obtained from Eq.(\ref{steady})) of two coupled SRR under the continuous tuning of the overall intensities of the two electromotive forces which deviate from each other by a small amount ($10^{-3}$). Although in the initial state the responses of the two SRR only have a tiny difference, the resulting states in the symmetry-breaking range can have a large contrast ratio. This property can be used to compare which signal (in our case the electromotive force) has a larger initial amplitude. Due to the small size of SRR compared to the working wavelength, the two SRR should be excited and detected via near-field ways (e.g. [\onlinecite{shadrivov2006oe}]). Such a device, if the two SRR can be fabricated as identical as possible, can be used to realize ultra-sensitive and possibly utlra-compact all-optical comparators in applied areas. We should stress that such applications do not have to rely on SRR, and they may be even realized in any other devices if two identical nonlinear resonators are strongly coupled with each other, resulting possible spontaneous symmetry breaking (e.g., [\onlinecite{davoyan2011,salgueiro2010,bulgakov2011}]). These devices do not involve any assistance from electronic devices and thus may be of interest in the fields such as all-optical logic gates, all-optical signal processing and photonic computers (e.g., [\onlinecite{mccutcheon2009,lu2011,wei2011}]). One remarkable feature of our proposal is that we do not need high quality factor as other applications for optical signal processing \cite{mccutcheon2009,tanabe2005,kim2007,husko2009,hu2008}, since we only assume $Q_f=10$ in our calculations. We also do not need uniform bistability.

\section{Setting metamaterials into Fabry-Perot (FP) cavities}\label{FP}

It is fascinating to see that due to field enhancement at the ``single-atom'' level, the nonlinearity of a single SRR can be strongly enhanced, and bistability can even occur for a single SRR at relatively low intensities, which certainly is an advantage of nonlinear metamaterials compared to natural materials. If SRR are simply stacked together, however, problems will arise as we have studied. One way out to utilize the enhanced nonlinearity is to focus on few-body devices as we studied in the previous section, i.e, directly making use of the multistabilities and modulational instabilities. Another is to operate the stacked devices under the low-intensity radiation and introduce \emph{global feed-back mechanism}, e.g., setting them (e.g., the structure in [\onlinecite{chen2011}]) in Fabry-Perot cavities with optical lengths much larger than the wavelength. This approach can not only effectively employ the enhanced nonlinearity originating from the sing-atom level, but also avoid the modulational instabilities suffered by the devices at higher intensities. That is because the FP-bistability requires a jump of the refraction index only of order $\lambda/L_{FP}$ (compared to the negative-positive jump proposed in [\onlinecite{zharov2003,chen2011}]), where $\lambda$ is the wavelength and $L_{FP}$ is the length of the cavity. Detailed estimations reveal that due to the field enhancement in the narrow gap, the required radiation intensity inside the FP cavity would be smaller by a factor of about $10^{-5}$ than that required if the FP is filled with the same kind of pure Kerr materials (see Appendix \ref{C}). This SRR-enhanced nonlinearity may be even more powerful and useful when combined with other nonlinearity-enhancing mechanisms, e.g., [\onlinecite{hosono1992,shimizua1992}].

\section{Discussions}

We want to comment on the relations between nonlinear metamaterials composed of SRR and natural nonlinear materials composed of natural atoms. The usually mentioned bistabilities in natural materials are induced by a global feed-back mechanism and work at a perturbative regime, i.e., the nonlinearity is restricted to the third-order. In other words, usual optical bistabilities are actually the nonlinear properties of certain global optical modes. No direct applications related to bistabilities in natural materials have been focusing on the non-perturbative regime in which the nonlinearity is very large at single-atom level or equivalently the radiation field intensity is comparable to the intrinsical field intensities in the atom level, since the responses may be very complex and very hard to control. Similarly, the proposals on controlling the bistabilities of the uniform optical constants \cite{zharov2003,chen2011} are just examples of working in the non-perturbative regime of the effective index of the bulk metamaterials, so it is not at all surprising to see the unpleasant instability problems. Actually, the uniform bistabilities in [\onlinecite{zharov2003,chen2011}] are properties of local resonance responses rather global modes in usual case. Our proposal to introduce global feed-back by setting an SRR array in an FP cavity is just an attempt to avoid these non-perturbative regimes and try to seek a perturbative regime (using optical modes) to utilize the SRR-enhanced nonlinearity. We illustrate the difference between bistabilities of local responses and global modes in Fig.\ref{unif}.

We also want to stress on the relations between the spatial modulational instabilities discussed in this paper and those for continuous nonlinear media. Continuous descriptions assume homogenization of a material and employ partial differential equations. The modulational instabilities and related solitons in continuous descriptions (e.g., [\onlinecite{wen2006}]), if indeed valid, are actually triggered by certain exploding long-wavelength (compared with the lattice constant) magneto-inductive waves \cite{shamonina2002} (or equivalents in other discrete systems) and at the same time preserved by the decaying short-wavelength magneto-inductive waves in the discrete descriptions. However, our results in Fig.1 and Fig.2 show no indication of such long-wavelength-only modulational instabilities even in the low intensity range. All the modulational-instability-induced nonuniform states in Fig.2 have short-wavelength components. This seems like a paradox.
Remember that the self-focusing-induced spatial modulational instabilities originate from the propagations of waves. As we state in the theoretical model section, we have only analyzed the modulational instabilities of an equiphase surface and have not involved the stabilities of propagations of waves, and therefore our results do not imply the paradox. These arguments implies that at comparatively low intensities, the spatial modulational instabilities can still happen due to propagation effects. As an added note, the modulational instability discussed in this paper should be termed as subwavelength modulational instability as in a recent publication [\onlinecite{noskov2012}].

 We should also note that in the perturbative range (or low intensity range) in Fig.\ref{unif}, the stable uniform responses enable the materials to be perturbatively characterized as nonlinear media with Kerr nonlinearity (also indicated in the Appendix \ref{C}). Therefore, the works (mentioned in the \emph{Introduction} section) which directly assume Kerr nonlinearities together with negative refraction indexes could be regarded as theories describing the perturbative properties of real discrete nonlinear metamaterials (e.g., as in \cite{lazarides2005}).

 One may question our negative results by referring to the experimental works \cite{shadrivov2008,powell2009}. Although frequency shift is demonstrated (certainly not sufficient for the index switching), no strong evidences (e.g., hysteresis curve) for bistability-related index switching are discovered, however. Moreover, the nonlinear metamaterials in [\onlinecite{shadrivov2008,powell2009}] are excited by monopole antennas, a nonuniform experiment condition way too far from explicitly demonstrating the uniform bistability of optical constants. In fact, the two papers only claim that their metamaterials are tunable and do not claim that the bistability is confirmed. Another question would arise from the LCR model we employ. Although this model is more or less simplified and is not general, it can capture the main features of many metamaterials, e.g., resonances, loss, mutual interactions, Kerr nonlinearity. Therefore, our results can give strong caveat for considering the applications of the nonperturbative behaviors of effective optical constants (e.g., bistability of refraction index) when interaction-induced modulational instability in the lattice level is possibly present, since the effective optical constants in those cases may be ill defined. We also want to point out that the recent publication \cite{noskov2012} mentioned above investigate another kind of nonlinear arrays. After making dipolar and slowly-varying approximation near a resonance, they derive almost the same formulism as in this paper and also obtain modulational instability in the uniform bistable range.

We also feel it is necessary to clarify the difference between the bistability of
optical constants in the proposals \cite{zharov2003,chen2011} and the bistability induced by periodic structures \cite{herbert1993,xu1993} in
the literatures. The origin of the former is the local
feed-back at the single-atom level while that of the latter is just the periodicity, (termed
as distributed feed-back in the seminal paper \cite{winful1979}). The
periodicity of the former has been homogenized, being
replaced by a homogenized optical constant. What is more, the former is derived by local
consistent equations of local optical constants and do not rely on the shape of the bulk
material (explicitly shown in [\onlinecite{zharov2003}]; in [\onlinecite{chen2011}] different shapes are
used with the same interpretation), while the latter is derived from the global consistent
equations explicitly determined by the boundary or periodicity. One can also introduce
such distributed or global feedbacks on a length scale larger than the wavelength to
achieve bistability in metamaterials as we propose in the previous section. These differences all arise from the working wavelengths of
these phenomena, one is larger than the lattice constant while the other is comparable or
smaller than that. Therefore, the historical success \cite{herbert1993,xu1993} cannot be evidences for the theoretical proposals in \cite{zharov2003,chen2011}.

Finally, we should be reminded that if the whole array of SRR are rather sparsely arranged, i.e., the coupling between neighbors can be neglected (gas phase), the local-bistability-based optical switching should also approximately be valid although the two switching points are inevitably affected according to Eq.(\ref{growrate}). But in those case the array is so sparse that it can hardly be treated as effective materials. The coupling parameters chosen in this paper is more typical in realistic cases (condensed phase).

\section{Summary}
By investigating the modulational instabilities and multistabilities, we find that some recently proposed optical switching based on the bistability arising from uniform assumptions may not be valid. In order to get rid of this dilemma, we propose two approaches to utilize the valuable enhanced nonlinearity of metamaterials at single-atom level, one is to focus on the multistabilities of few-body problems (e.g., all-optical comparator) and the other is to operate them at low radiation intensities by introducing global feed-back mechanism to achieve global bistability effects (e.g., Fabry-Perot cavity) in perturbative regimes of optical constants. Finally, we briefly discuss the relations between our results and existing ones in the literatures. Particularly, we distinguish the questioned bistability supported by local resonance and the usual bistability of global optical modes (global resonances, or global feed-backs) and stress that the former may easily be destroyed by modulational instability induced by discreteness and mutual interaction in dense materials.

\section*{Acknowledgement}
We especially thank Cuicui Lu and Junhui Li for their careful reviews of this paper and constructive suggestions. We also thank the whole of Xiaoyong Hu's group for helpful discussions and A. Alu for his comments and providing related literatures.

\appendix
\section{The Theoretical model}\label{A}
The typical structure of SRR is shown in Pendry's seminal paper \cite{pendry1999}.  According to the effective LCR-circuit model of the SRR in the static limit, we can write down the dynamic equation of SRR lattice.
\begin{eqnarray}
&&L\frac{d^2Q_{i,j}}{dt^2}+R\frac{dQ_{i,j}}{dt}+\frac{Q_{i,j}}{C}=Ue^{i\omega t}\nonumber\\
&&-M_x\left(\frac{d^2Q_{i+1,j}}{dt^2}+\frac{d^2Q_{i-1,j}}{dt^2}\right)\nonumber\\
&&-M_y\left(\frac{d^2Q_{i,j+1}}{dt^2}+\frac{d^2Q_{i,j-1}}{dt^2}\right)\label{1}
\end{eqnarray}
where $L$, $R$ and $C$ is the effective self-inductance, resistance, and capacitance, respectively. The $Q_{i,j}$ is the charge hold by the capacitor at the site $(i,j)$. The $M_{x,y}$ are mutual inductances between nearest neighbors in the x (y) directions, respectively. Note that the nearest-neighbor approximation is employed.

In our case where the split of the ring is filled by Kerr materials, the effective capacitance depends on the electric field intensity, $E_g$, in the gap, i.e., $C=C_0(1+|E_g|^2/|E_c|^2)$. Here, the $C_0$ is the zero-field capacitance and the $|E_c|$ is the characteristic field intensity.

We also have the relation between $Q$ and $E_g$ in the quasi-steady states,
\begin{equation}
Q=CE_gd_g.
\end{equation}
where $d_g$ is the gap width of the capacitor. Then we arrive at the relation between $Q$ and $C$,
\begin{equation}
C=C_0\left(1+\frac{|Q|^2}{C^2d_g^2|E_c|^2}\right).
\end{equation}
Here since we usually work in the weak-nonlinearity regime of the Kerr material, the above relation can be approximated as,
\begin{equation}
C\approx C_0\left(1+\frac{|Q|^2}{C_0^2d_g^2|E_c|^2}\right).
\end{equation}
and
\begin{equation}
\frac{1}{C}\approx \frac{1}{C_0}\left(1-\frac{|Q|^2}{C_0^2d_g^2|E_c|^2}\right).\label{2}
\end{equation}
Combining the Eqs. (\ref{1}, \ref{2}) and make slowly-varying-amplitude approximation, we can obtain the dynamic equation (\ref{dynamic}) after defining the following dimensionless quantities. $q=\frac{Q}{C_0d_g|E_c|}$, $\omega_0=1/\sqrt{LC_0}$, $\Omega=\omega/\omega_0$, $\gamma=RC_0\omega_0$, $\kappa_{x,y}=M_{x,y}/L$, $u=U/(|E_c|d_g)$ and $e=E_g/|E_c|=q$.

For the convenience of the discussion in Sec.(\ref{FP}), we write down the nonlinear magnetic susceptibility of a single SRR.
\begin{equation}
\chi\propto\frac{I}{U}\propto\frac{1}{Z}.
\end{equation}
where $I$ is the current and $Z$ is the complex impedance.
With the linear part $\chi_l$ separated, it reads
\begin{equation}
\chi=\chi_l\frac{-\Omega^2+i\gamma\Omega+1}{-\Omega^2+i\gamma\Omega+1-|e|^2}
\end{equation}

\section{Linear stability analysis}
We study the linear stability of uniform states $q_{i,j}=q$ solved from the steady state equation (\ref{uniform-steady}). We first introduce tiny fluctuations to the steady states, $q_{i,j}=q+\Delta_{i,j}$, and derive the linear dynamic equations for $\Delta_{i,j}$ by use of the full dynamic equation.

\begin{eqnarray}
&&(2i\Omega+\gamma)\frac{d\Delta_{m,n}}{d\tau}+2i\Omega\kappa_x(\frac{d\Delta_{m-1,n}}{d\tau}\nonumber\\
&&+\frac{d\Delta_{m+1,n}}{d\tau})+2i\Omega\kappa_y(\frac{d\Delta_{m,n-1}}{d\tau}+\frac{d\Delta_{m,n+1}}{d\tau})\nonumber\\
&&=\kappa_x\Omega^2(\Delta_{m-1,n}+\Delta_{m+1,n})+\kappa_y\Omega^2(\Delta_{m,n-1}\nonumber\\
&&+\Delta_{m,n+1})-(-\Omega^2+i\gamma\Omega+1-2|q|^2)\Delta_{m,n}\nonumber\\
&&+q^2\Delta^*_{m,n}
\end{eqnarray}

Since the lattice is uniformly periodic, the eigenmodes should be uniform plane waves. Thus let $\Delta_{m,n}(\mathbf{k})=\alpha e^{ik_xm+ik_yn}+\beta e^{-ik_xm-ik_yn}$. We then have
\begin{eqnarray}
&&\frac{d}{dt}\left(\begin{array}{c}\alpha\\ \beta^*\end{array}\right)=\frac{1}{i\eta}\left(\begin{array}{cc}A&B\\ -B^*&-A^*\end{array}\right)\left(\begin{array}{c}\alpha\\ \beta^*\end{array}\right)\nonumber\\
\end{eqnarray}
where $A=2\kappa_x\Omega^2\cos k_x+2\kappa_y\Omega^2\cos k_y-(-\Omega^2+i\gamma\Omega+1-2|q|^2)$, $B=q^2$ and $i\eta=\gamma+2i\Omega(1+2\kappa_x\cos k_x+2\kappa_y\cos k_y)$.The eigenvalues $\lambda$ of the matrix on the left are the growth rates of the magneto-inductive fluctuations with wave vector $(k_x,\,\,k_y)$.

\begin{eqnarray}
&&\lambda(\mathbf{k})=\frac{1}{\eta}[-\gamma\Omega\pm(|q|^4-(2\kappa_x\Omega^2\cos k_x+\nonumber\\
&&2\kappa_y\Omega^2\cos k_y+\Omega^2-1+2|q|^2)^2)^{1/2}]\label{growrate}
\end{eqnarray}

\section{Estimations in Sec.\ref{FP} }\label{C}
In the Sec.\ref{FP}, we propose bistability by setting an SRR array into an FP cavity. First of all, for the bistability to occur the required radiation intensity should reach a certain scale to induce nonlinear deviation of the refraction index $\Delta n\sim \lambda/L_{FP}$, where $\lambda$ is the radiation wavelength in the metamaterial and $L_{FP}$ is the length of the FP cavity. This is usually a small factor provided that $L_{FP}\gg\lambda$, which will effectively reduced the required radiation intensity and thus help to avoid the modulational instabilities at higher radiation intensities. In this regime, the nonlinearity of the metamaterial can be treated perturbatively.

Thus we need that the radiation propagate at least $L_{FP}$ distance before it significantly attenuate.
To this end, we should work at the frequencies where the metamaterial is transparent, i.e., far from the absorption ($|\delta|=10/Q_f$, where the detuning $\delta=(\omega-\omega_0)/\omega_0$), and therefore the lossless approximation $R=0$ can be justified.  According to the results in the Theoretical Model section, the effective magnetization coefficient of the metamaterials can be approximately written in this case as $\chi(|e|)=\chi_l[1-|e|^2/(2\delta)]$. To express $\chi$ in term of the radiation intensity $|E|^2$, we should relate the dimensionless field intensity $|e|$ in the gap to the radiation intensity $|E|$.

The total electromotive force accumulated along the ring is
\begin{equation}
|U|=|\int(i\bf{k}\times\bf{E})\cdot d\bf{S}|\sim \frac{2\pi^2r^2}{\lambda}|E|,\label{EMF}
\end{equation}
with $r$ being the radius of the SRR.
Then in the lossless limit,the voltage $V$ across the gap will be
\begin{eqnarray}
V&=&\frac{U}{i\omega L+1/(i\omega C)}\cdot \frac{1}{i\omega C}\nonumber\\
&\approx& \frac{U}{1-\omega^2/\omega_0^2}\nonumber\\
&\approx&-\frac{U}{2\delta}.\label{enhancement}
\end{eqnarray}

Now, considering the relations $E_g=V/d_g$ and $e=E_g/|E_c|$, we have by relating the above results,
$$|e|=\frac{\pi^2r^2}{|\delta|\lambda d_g}\frac{|E|}{|E_c|}.$$
Then $\chi(|E|)=\chi_l[1-\frac{\pi^4r^4}{2\delta^3\lambda^2d_g^2}\frac{|E|^2}{|E_c|^2}]$, which indicates that self-focusing and self-defocusing nonlinearity are both possible in the perturbative regime depending on the sign of the detuning $\delta$ and $\chi_l$.

Now, we can estimate the radiation intensity required to induce the bistability of an FP cavity,
$$\chi_l\frac{\pi^4r^4}{2|\delta|^3\lambda^2d_g^2}\frac{|E|^2}{|E_c|^2}\sim \frac{\lambda}{L_{FP}},$$
(for the bistability without FP cavity, the right-hand side will be of $\mathcal{O}(1)$.)then
$$|E|^2/|E_c|^2=\frac{\lambda}{L_{FP}}\frac{2|\delta|^3\lambda^2d_g^2}{\pi^4 r^4\chi_l}.$$
Compared with an FP cavity filled with the same kind of pure Kerr material, the required intensity inside the cavity is reduced by a factor
\begin{equation}
\frac{2|\delta|^3\lambda^2d_g^2}{\pi^4 r^4\chi_l}.\label{factor}
\end{equation}

For the structure proposed in \cite{chen2011}, if the $Q_f$ can reach 100 (see Fig.3a of \cite{chen2011}), we can estimate $\delta=0.1$. The expression for electromotive force (\ref{EMF}) and the effective-$\chi$ description should be still roughly valid for a wavelength-size ratio $\lambda/r$ as small as 5. The gap-size ratio $d_g/r$ is a tunable parameter depending on the fabrication technique which we assume can reach as small as 1/10. We also assume that the linear response (depending on the detuning and density of SRR) is not too small, i.e., $\chi_l\sim\mathcal{O}(1)$. Then the reduction factor could be estimated to be about $10^{-5}$.

The role of the resonance can be assessed, from the above estimations, to be positive because it can render large field enhancement [$\delta$ should better be small in Eq.(\ref{enhancement})]; negative because it inevitably bring in energy loss that may make the material opaque ($\delta$ should not be too small).


%

\end{document}